# An authorship protection technology for electronic documents based on image watermarking

**Anna Melman**[a], **Oleg Evsutin**[b], **Alexander Shelupanov**[a]

[a] Tomsk State University of Control Systems and Radioelectronics, 40 Lenina Prospect, Tomsk 634050, Russia

[b] National Research University Higher School of Economics, 20 Myasnitskaya Ulitsa, Moscow 101000, Russia

Correspondence: Oleg Evsutin, evsutin.oo@gmail.com

**Abstract**

In the field of information technology, information security technologies hold a special place. They ensure the security of the use of information technology. One of the urgent tasks is the protection of electronic documents during their transfer in information systems. This paper proposes a technology for protecting electronic documents containing digital images. The main idea is that the electronic document authorship protection can be implemented by digital watermark embedding in the images that are contained in this document. The paper considers three cases of using the proposed technology: full copying of an electronic document, copying of images contained in the document, and copying of text. It is shown that in all three cases the authorship confirmation can be successfully implemented. Computational experiments are conducted with robust watermarking algorithms that can be used within the technology. A scenario of technology implementation is proposed, which provides for the joint use of different class algorithms.

**1. Introduction.**

The widespread adoption of information technology in all spheres of life leads to the fact that paper media are gradually losing their popularity. This means that the number of documents submitted in electronic form is constantly increasing. Many private and public institutions use electronic workflows. A large number of people refuse to buy hard-copy books, newspapers and magazines in favor of their electronic versions. Various advertising materials, posters, announcements are also successfully distributed in electronic form. It is highly convenient for users, but at the same time it is highly convenient for adversaries, because they can easily appropriate other people's intellectual output. Therefore, at present, the problem of protecting the authorship of electronic documents is relevant and requires close attention.

The main way to protect the authorship of electronic documents is the use of a digital signature. The digital signature is an alternative to a handwritten signature and is used when organizing a secure electronic document flow. The digital signature allows you to confirm the fact of making changes to an electronic document after it is signed. However, it does not allow you to establish the fact of illegal copying of information, including digital images, into other documents. In some cases, it may also be necessary to protect individual document fragments. The digital signature can be applied to individual fragments of an electronic document, but the simultaneous transmission and storage of a large number of digital signatures causes inconvenience.

Along with the use of cryptographic methods for data protection, it is effective to embed additional data into digital covers, mainly into multimedia content. There are two directions for data hiding in digital objects: steganography and watermarking.

Steganography aims to protect the confidentiality of information. Protected data is embedded in some cover object to become invisible for third parties. Steganography is an alternative to encryption. Encryption makes information unreadable in the absence of a secret key. However, encryption does not hide the fact that protected information exists. Steganography conceals the existence of a secret message.

Digital watermarks are used to protect the authorship or integrity of the cover object itself. The extraction of embedded information allows you to confirm, for example, that a digital object was created by a specific person or software. To do this, the extracted watermark is usually compared to the original watermark. The extracted watermark and the original watermark are expected to match each other. The match can be complete or partial. It depends on the specific use case for digital watermarks.

The main idea of our research is to protect electronic documents by embedding digital watermarks in digital images contained in these documents. Unlike text, which is often revise when copied, images are usually copied unaltered, since reproduction of such an image can be time consuming or even impossible. The copied image processing does not destroy the embedded watermark. This allows you to successfully prove authorship if necessary.

It is obvious that the protected electronic documents must contain some kind of graphic object to implement this approach. However, at present this is not a problem, since many electronic documents are accompanied by various illustrations, diagrams, photographs, logos and other graphic elements.

Thus, the paper proposes a technology for protecting the authorship of electronic documents containing images through the use of digital watermarks. The limits of applicability of the proposed technology are investigated using existing watermarking algorithms of different classes.

The rest of the paper is organized as follows. Section 2 contains the literature review. In section 3, we propose the technology for electronic documents authorship protection and we also describe its application scenarios. Section 4 shows the results of computing experiments with three different robust watermarking algorithms. Section 5 contains a discussion of the results. Section 6 sums up our research.

## 2. Related work

The authorship protection of electronic documents without the use of cryptographic techniques investigated by many authors, but a universal solution does not currently exist. An actively developing area of protection of the authorship to electronic documents is the embedding of digital watermarks into documents. Watermarks can either be generated based on the protected text itself, or they can be text independent and contain some information about the author.

Text watermarking methods are divided into two large classes: linguistic and structural [1].

Linguistic methods are based on changing the syntactic or semantic structure of the text. For example, the authors of [2] propose a method for hiding watermark bits by lexical or syntactic changes of texts in German. The changes are based on grammar rules related to letter doubling, adjective formation, etc. The paper [3] presents a method for digital watermarks embedding in texts in English. The authors use the grammatical rules of the language and the most commonly used words such as conjunctions, pronouns and modal verbs to form watermarks. These watermarks are then embedded in web documents.

Such methods are resistant to text formatting as well as retyping attack. However, embedded watermarks are destroyed when text processing, for example, by replacing the word order or using synonyms. Another disadvantage is changes of the protected text, because this is unacceptable in some cases.

Structural methods deal with the formatting of the text. They perform various manipulations with the appearance and position of text characters. For example, the authors of [4] propose to embed a watermark based on biometric characteristics into a document by shifting lines relative to each other. The paper [5] describes a method for protecting text in Arabic, which includes modifying the lengths of spaces and using a special character of the language. At the first stage of the algorithm presented in [6], a watermark is generated based on the user's password and the original text using a hash function, and at the second stage, the created watermark is embedded into the text using homoglyph characters. In [7], it is proposed to change the line spacing when embedding a watermark to ensure resistance to printing and scanning attack. A general lack of such methods is that the text formatting can be easily changed using any modern text editor. As a result of editing, the embedded information is lost.

The popularity and widespread use of the PDF format for storing electronic documents attract the attention of watermarking method researchers. A common approach to digital watermarks embedding in PDF documents is character position manipulation. Some embedding schemes change the length of spaces between words, others change the length of spaces between characters within words. The watermarking scheme presented in [8] is based on the first method. Also, its distinctive feature is that it is not the length of spaces that change, but the frequency spectrum of the corresponding vectors, obtained using the discrete cosine transform (DCT). Scheme [9] is based on the second method. The authors point out that in PDF documents, each character has a coordinate pair, x and y, that locates the character horizontally and vertically within two-dimensional coordinate space. The authors propose to change x-coordinates when embedding a digital watermark. The development of this scheme is presented in a later work [10]. These methods allow you to protect a PDF file from unauthorized distribution. However, when copying text content, all information about character positions is lost, and embedded digital watermarks are useless.

A special class of methods is embedding information into text images. Text images are images that contain text, such as scanned copies of paper documents. [11] proposes a cloud-based approach to digital watermarks embedding in text images. Another example of text image watermarking is presented in [12], where the watermark is embedded in pixels using the linear interpolation technique. The embedding method described in paper [13] combines two frequency transforms: integer wavelet transform (IWT) and DCT. The authors demonstrate the effectiveness of this method using the example of images containing text in Arabic. The authors of the study [14] propose to change the boundaries of text characters in an imperceptible way that allows you to ensure resistance to printing and scanning.

In [15], the authors propose to embed additional information into images contained in Microsoft Word documents. The authors propose to hide secret messages in this way for their subsequent transmission. However, embedding is not used to protect the authorship of the document, but to ensure the confidentiality of the embedded information. The resistance of embedding to any typical transformations is not investigated.

It should be noted that research in the field of image data hiding is actively developing in recent years. Steganographic embedding provides an invisible transfer of information within the image [16-18]. Watermarks are designed to control the integrity, authenticity, and authorship protection of images. Watermark embedding in digital images contained in electronic documents uses the same techniques as embedding watermarks in individual images. We also give a brief overview of the current state of the relevant research area.

Digital watermarks differ in the level of resistance to distortion. Fragile watermarks are destroyed whenever a cover image is changed. Semi-fragile watermarks can withstand some attacks, such as moderate JPEG compression. Robust watermarks are also detectable after more significant distortion of the cover image. A digital watermark is usually some kind of user information such as a logo or text. Or it is a bit sequence of limited size, generated based on user data or the cover image content. Embedding a large amount of information as watermarks is not a promising area of research, since it usually does not allow the development of robust embedding algorithms. Therefore, a number of papers are presented here that describe the embedding of small watermarks.

Image watermarking methods are divided into spatial and frequency domain methods. Spatial domain methods work mainly with the pixel values of images. An example of a recent spatial domain method with a declared high efficiency is [19]. The authors of this study propose embedding a watermark by quantizing the pixels of a digital image. A binary image is used as a watermark. Arnold transform is applied to it at preprocessing stage to improve security. An embedding path is determined by hash pseudorandom scrambling algorithm. A feature of the algorithm is calculating DC coefficients of two-dimensional discrete Fourier transform (DFT) without performing a real 2D-DFT to reduce the running time of the algorithm.

Frequency domain embedding uses different frequency transformations, for example, DCT, DFT, discrete wavelet transform (DWT) and others. Many methods, especially early ones, use DFT, for example [20]. There, a specially generated circular watermark is embedded in the amplitude spectrum in an additive and multiplicative manner. In [21], the watermark bits are first embedded in the DFT amplitude spectrum, then the 2D histogram of the chromatic components Cb and Cr is modified.

In recent years, most researchers deal with DCT and DWT. For example, the authors of [22] hide a watermark in the mid-frequency DCT coefficients of digital images. Authors embed watermark bits in 8x8 blocks. One bit of the watermark is embedded in one image block by changing the difference between adjacent blocks. The authors divide the possible values of the difference into intervals that correspond to 0 or 1. They change the frequency coefficients so that their difference falls within the appropriate interval.

An example of frequency domain embedding algorithm based on DWT is described in paper [23]. A key feature of this study is the use of a differential evolution algorithm to find the best location for watermark blocks.

In some studies, authors use transformation compositions. An example of such an approach is described in [24]. The authors use a hybrid embedding scheme that combines DCT and DWT. First, DCT is applied to the pixels of the cover image. Then the Haar DWT is applied to the obtained values to obtain four frequency sub-bands. Fragments of the digital watermark are embedded to these 4 sub-bands in an additive manner after the preliminary application of Arnold transform and DCT.

There are also algorithms based on other less common transformations. For example, in [25], the contourlet transform and the Fresnel transform are used. A QR code is used as a watermark, and the optimal frequency coefficients for embedding are selected using optimization algorithms. The authors of [26] use the Walsh-Hadamard transform to embed watermarks in digital images. The embedding algorithm is developed using a linear prediction function.

Thus, various authors develop many different watermarking methods. In particular, there are many methods for hiding watermarks in digital images. Many of them are claimed to be quite effective. We have selected examples of different class methods to study their applicability within the proposed technology. The proposed technology and experimental results are presented below.

## 3. The proposed technology

In this section, we propose the electronic documents authorship protection technology based on image watermarking. We discuss scenarios for using this technology and options for its implementation.

*3.1. The main concept*

Section 2 presented examples of various methods for digital watermark embedding in electronic documents. Each class of these methods has both advantages and disadvantages that limit their practical applicability.

We propose a technology for protecting electronic documents based on hiding digital watermarks in images contained in documents. Unlike the methods presented in Section 2, our technology does not imply any changes to the text or its formatting. Instead, invisible watermarks are embedded in graphics contained in an electronic document.

Before inserting images into the document created in a text editor, the content author must embed watermarks (the same or different) into all these images according to the chosen embedding algorithm. It should be especially noted that to ensure a high level of protection, robust watermarks should be used that are not destroyed under various destructive effects on the image. After placing watermarked images in the document, this document can be stored or transmitted in its original form or after conversion to another format, such as PDF. The general scheme of this process is shown in Fig. 1.

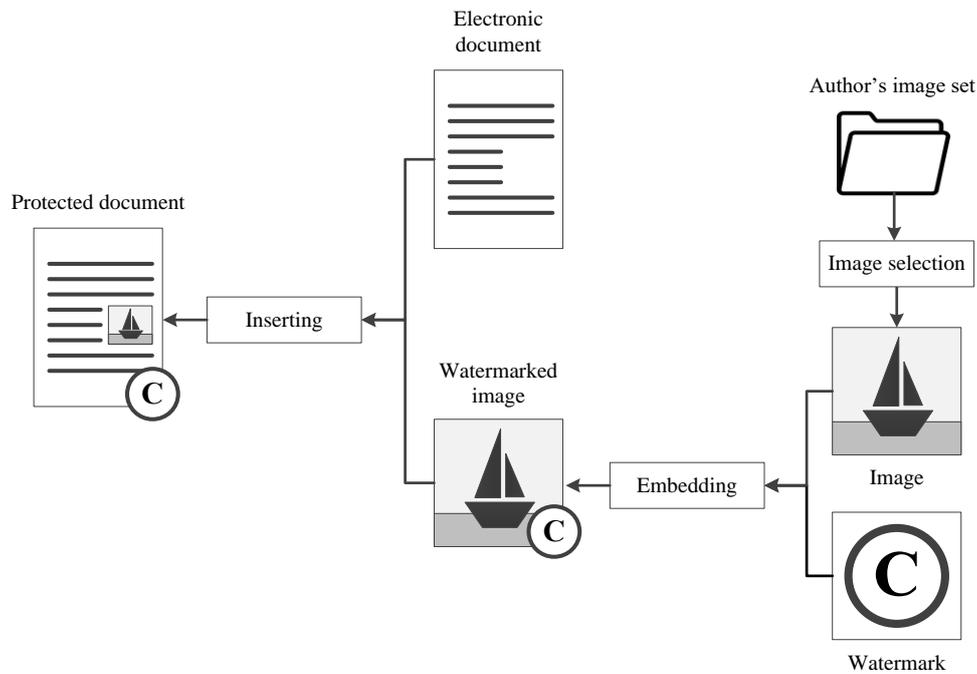

**Fig. 1.** General scheme of document protection according to the proposed technology

The proposed protection technology allows us to implement an electronic document authorship protection without changing the content and structure of the text. A well-chosen watermarking algorithm ensures a high level of robustness. Copied images will still have been watermarked even after distortions such as compression, brightness change, cropping, scaling, etc.

Obviously, the proposed technology has a limitation, because the protected electronic document must contain a graphic object. However, despite this limitation, this technology can be used to protect a large number of electronic documents. Nowadays, many documents contain some kind of graphic elements. Even corporate documents often contain a logo, which can be used as a cover image for a watermark. We also note that the problem of authorship protection for digital content is more typical for electronic documents related to creative and intellectual work. The results of such work are books, papers, blog posts and other publications, which are usually accompanied by diagrams and illustrations. Therefore, the target audience of the proposed technology is mainly the authors of original content in different spheres.

Unlike software that completely prohibits copying information, our technology allows you to copy text and graphic objects from a document. We mean the case when the author publishes a document in the public domain and does not object to the use of its fragments, if such use does not infringe copyright. Regardless of the purpose for which someone copies a document, an embedded watermark can successfully identify the original author of the content.

The author of the content can use his identification data and brief information about the protected document as a digital watermark. For the most effective authorship protection of an electronic document, it is necessary to ensure the link between the image and its context in the document. In this case, to generate a watermark, the identification data of the author, the name of the document and a fragment of its text are required. The association of a digital watermark with the text of a document can be used to prove authorship even if the text of the protected document is copied without a graphic object.

The case when the watermark is specified by the user and does not depend on the context of the image in the document is illustrated in Fig. 2a. In this case, the author of digital content chooses some information that is used as a watermark. It can be a text about the author or a personal logo. This data can be used as a watermark without any changes or after applying additional transformations. The author embeds the selected watermark into the image, which is then placed in the document.

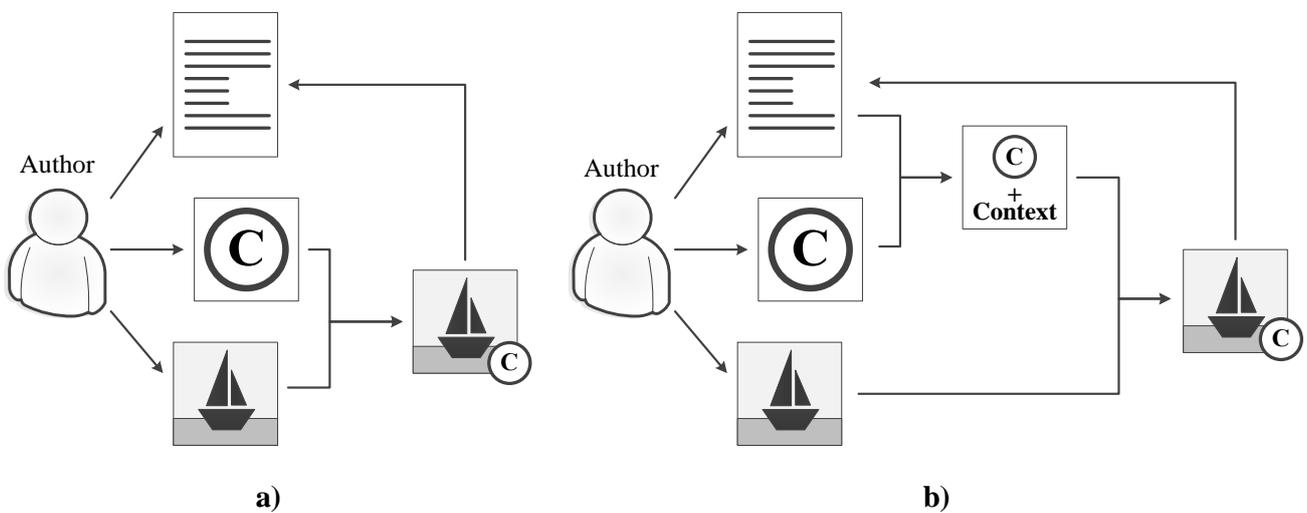

**Fig. 2.** Embedding scheme: **a)** The watermark is not based on the image context; **b)** The watermark is based on the image contex

This scheme hardly differs from the classic case of embedding digital watermarks into digital images, especially if the resulting document is stored in its original format. The situation when a document is converted into any graphic format or into PDF format is of greater interest. In this case, the images contained in the document are distorted. Therefore, to protect authorship, it is necessary to use robust watermarks, in particular, resistant to image compression.

The advantage of using this embedding scheme is user friendliness. No additional steps are required other than choosing a watermark for embedding. However, this scheme has a significant drawback. If a plagiarist copies the text of a document without images, the link between the images and the document is lost. Subsequently, it is difficult to prove authorship for the entire document, and not just for an individual image.

In another possible scheme, the watermark is generated based on user information as well as the context of the image. Image context is the text fragments or other objects surrounding the image in an electronic document or information about the location of the image (page number, paragraph, line, etc.). In this case, the step of watermark embedding is preceded by the step of watermark generating. User information is combined with context information, for example, by concatenation. Hashing can be used to obtain a watermark of a fixed small size. The described embedding scheme is illustrated in Fig. 2b.

This embedding scheme is more difficult for the user, compared to the previous one, because it is necessary to set the context for each specific image. At the same time, the use of the context significantly increases the security of the proposed technology. Even if the text of the document is copied without images, the author can easily prove the authorship by demonstrating the link of the digital watermark with the text. The use of context is useful for authorship proof in a situation that the text copied by a plagiarist is significantly processed. The watermark shows a link to the original document even after text changes. This variant of the proposed technology application is recommended for practical use.

Note that the proposed technology can be used in combination with other technologies for protecting electronic documents, if an increased level of security is required. The combination of this technology with text watermarking methods provides reliable protection of authorship for an electronic document at several levels at once.

*3.2. Application scenarios*

In practice, when using the proposed technology, three main scenarios can be carried out. The application scenario is the situation of copying an electronic document, when authorship proof becomes relevant. The purpose of the copy is not analyzed. In terms of the proposed technology, it does not matter who the plagiarist is and why he or she copies information. The main scenarios include copying of an entire electronic document, as well as partial copying. Partial document copying can be divided into image copying and text copying. Below, we discuss each scenario to analyze the effectiveness of the proposed technology.

*3.2.1. Complete copying*

Complete copying of a document means copying the entire content of an electronic document, including text and images. This is the most likely scenario when text and images are logically related and the meaning of the document is distorted in the absence of text or graphics. In terms of the proposed technology, it does not matter whether the plagiarist copies the document file or copy its contents to another file.

Since watermarked images were copied along with the text of the document, you only need to extract the watermarks from them to prove authorship. If the author of the original content uses a robust watermarking algorithm, proof of authorship is successful even after some typical image processing techniques are applied to the graphic objects. Revising the text does not affect the effectiveness of proof of authorship as watermarks are embedded in images.

In this scenario, a high level of the proposed technology efficiency can be achieved both by embedding an independent watermark and a context-based watermark. The described scenario is illustrated in Fig. 3.

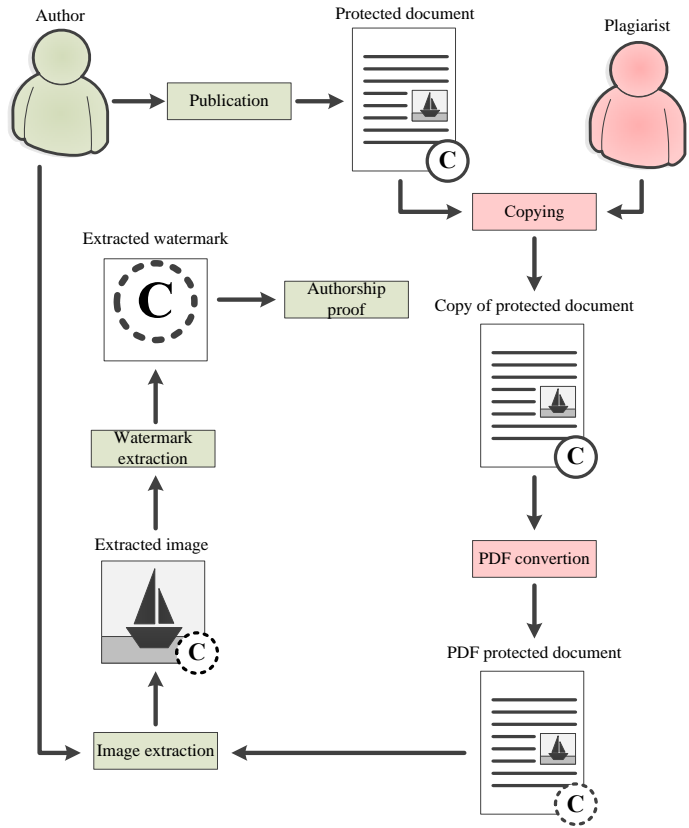

**Fig. 3.** Authorship proof scenario when complete document copying

*3.2.2. Text only copying*

The plagiarist can copy the text of a protected document only. This scenario is most likely for electronic documents for which images are mostly decorative. Since the text of an electronic document is not protected by a watermark, the authorship proof procedure requires the original version of an electronic document containing watermarked graphic objects. If proof of authorship is required, the author of the original content must provide the original version of the electronic document and extract the watermarks from the images. In this case, the high efficiency of the proposed technology is ensured if the image context in the document is used when generating the watermark. The robustness of the watermark is optional. The described scenario is illustrated in Fig. 4.

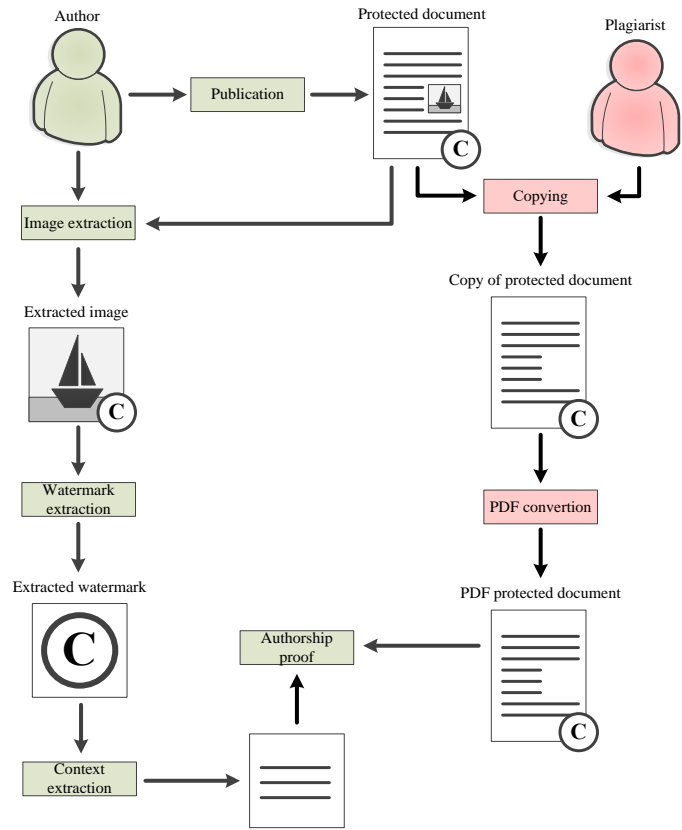

**Fig. 4.** Authorship proof scenario when text only copying

### 3.2.3. Images only copying

The plagiarist can copy only graphic objects (photographs, illustrations, diagrams, etc.) into another document. He or she can also copy these graphic objects without placing them in another electronic document. This is the most likely scenario if the document consists mainly of images. This case is completely analogous to the classic scenario of digital watermark embedding into images. The invisible watermark must be extracted from the image to prove authorship. If the author of the original content uses a robust watermarking algorithm, authorship proof is successfully implemented even after distorting the watermarked image. The effectiveness of the proposed technology is high whether the author uses the independent watermark or the context-based watermark. This scenario is illustrated in Figure 5.

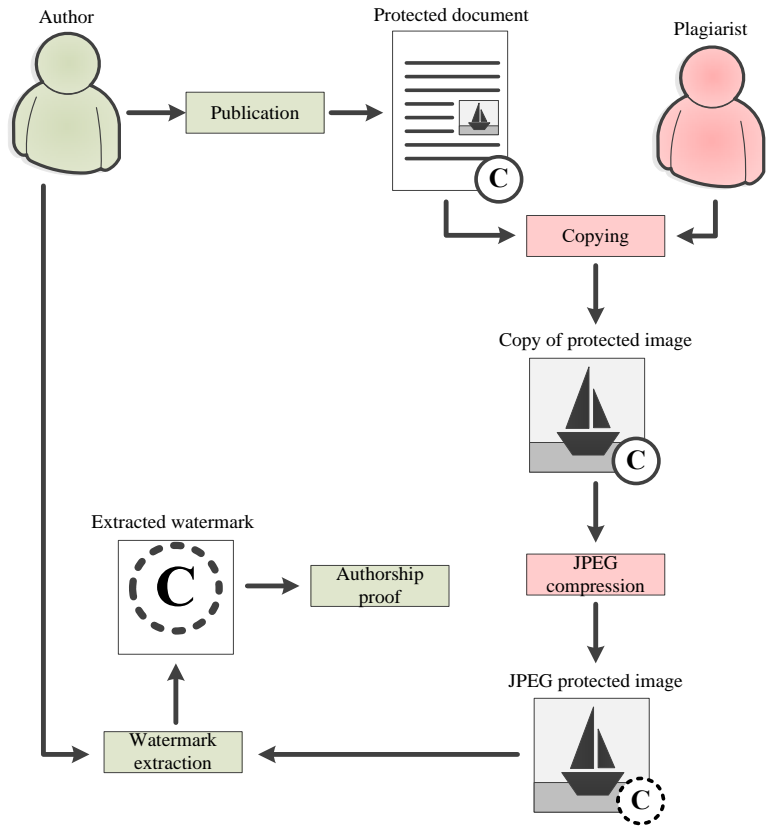

**Fig. 5.** Authorship proof scenario when images only copying

In all three scenarios, the authorship protection can be successfully implemented. It is recommended to use robust watermarking algorithms and linking to the electronic document context to ensure a high level of authorship protection.

## 4. Experimental results

To carry out the experiments, we create electronic documents containing text and halftone 512 × 512 images using Microsoft Word text editor. We use 10 images in our experiments. Half of the images are from The USC-SIPI database [27]. These are the classic images such as "Airplane", "Baboon", "Barbara", "Goldhill", and "Peppers". Five more images are the stock images from pixabay.com [28]. These are illustrations containing a large number of smooth areas. This set of images is due to the fact that electronic documents can contain images of different types. These can be photographs containing many objects and textures, as well as various drawings and schemes. The cover images are shown in Fig. 6.

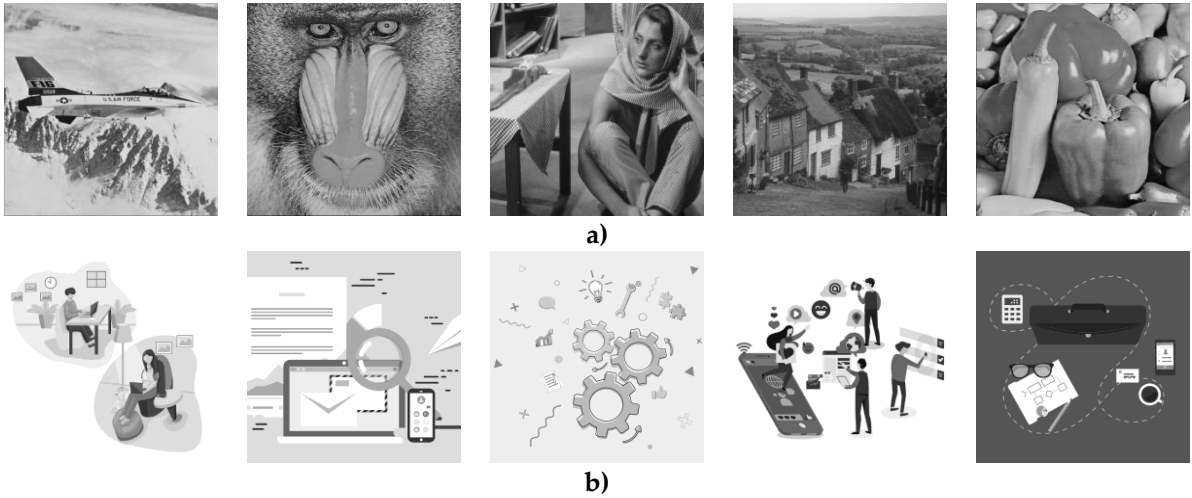

**Fig. 6.** Cover images: (a) from The USC-SIPI database; (b) from pixabay.com

A 64 × 64 binary watermark is embedded in each of the images. In these experiments, we do not use the context for watermark generation. The purpose of the experiment is to test the resistance of the watermarked images to the document file conversion into PDF format. We chose to convert document file to PDF because it is one of the most common conversions for electronic documents.

We use the algorithms described in papers [19], [22] and [24] for embedding. These are robust watermarking algorithms of recent years, characterized by increased resistance to JPEG compression. JPEG compression is most commonly used when converting documents to PDF. Algorithm [19] is an example of spatial domain embedding, and [22] and [24] are examples of frequency domain embedding. A more detailed description of these works is given in Section 2. Watermarked images are shown in Fig. 7.

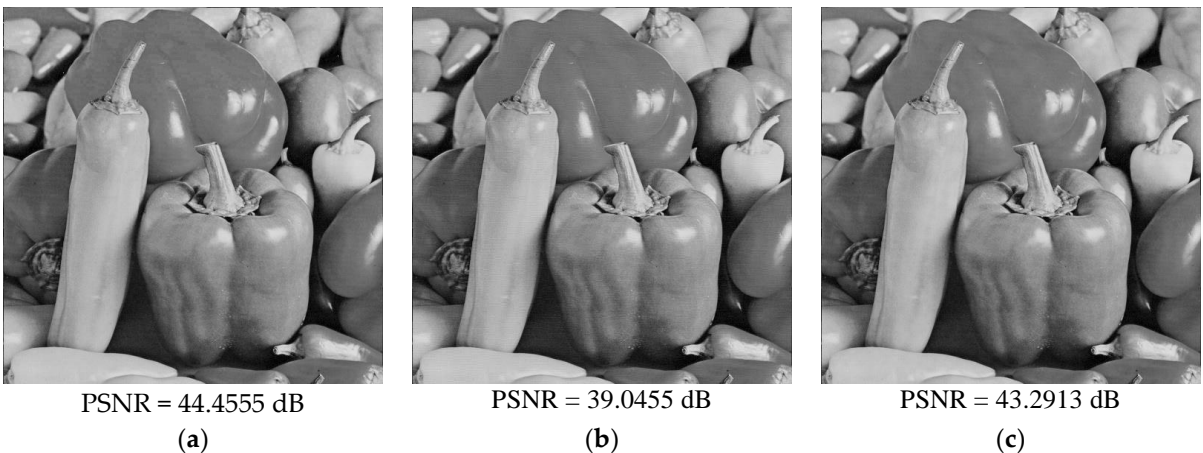

PSNR = 44.4555 dB     PSNR = 39.0455 dB     PSNR = 43.2913 dB
(a)                   (b)                   (c)

**Fig. 7**. Watermarked images obtained using: **a)** algorithm [19]; **b)** algorithm [22]; **c)** algorithm [24].

The resulting electronic documents are converted to PDF files with various graphics compression quality settings. Next, watermarked images are extracted from the PDF document, and watermarks are extracted from images.

The bit error rate (BER) and normalized cross-correlation (NCC) are used to assess the quality of the extracted watermarks. BER can be expressed by the formula

$$\text{BER} = \frac{B_e}{B} \qquad (1)$$

where $B$ is the number of watermark bits, $B_e$ is the number of errors (changed bits) that occurred during extraction.

NCC between the original watermark $W$ and extracted watermark $W_{ext}$ is computed using the formula

$$\text{NCC} = \frac{\sum_{x=1}^{M}\sum_{y=1}^{N}(W(x,y) \times W_{ext}(x,y))}{\sqrt{\sum_{x=1}^{M}\sum_{y=1}^{N}(W^2(x,y))}\sqrt{\sum_{x=1}^{M}\sum_{y=1}^{N}(W_{ext}^2(x,y))}} \qquad (2)$$

where $M \times N$ is a size of original watermark.

Tables 1 and 2 show the results of watermark extraction from images obtained by the spatial domain embedding algorithm [19]. The tables show the compression quality, BER and NCC values, and demonstrate the extracted watermark.

The experimental results confirm the high resistance of the algorithm [19] to image compression. The watermark is extracted in its original form or with minor distortion after maximum and high quality compression. Fragments of the watermark can be found in the protected image even after minimum quality compression. Watermark embedding in the illustrations shows higher resistance to compression than watermark embedding in highly detailed images. High robustness of embedding makes it possible to effectively protect the authorship of electronic documents containing images.

**Table 1.** Experimental results for highly detailed images watermarked by the algorithm [19]

| Compression quality | Airplane | Baboon | Barbara | Goldhill | Peppers |
|---|---|---|---|---|---|
| No | 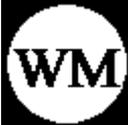<br>BER = 0<br>NCC = 1 | 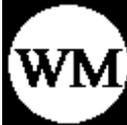<br>BER = 0<br>NCC = 1 | 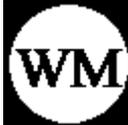<br>BER = 0<br>NCC = 1 | 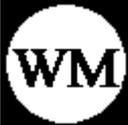<br>BER = 0<br>NCC = 1 | 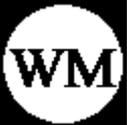<br>BER = 0<br>NCC = 1 |
| Maximum | 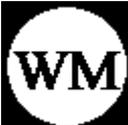<br>BER = 0<br>NCC = 1 | 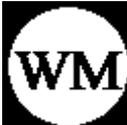<br>BER = 0<br>NCC = 1 | 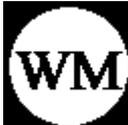<br>BER = 0<br>NCC = 1 | 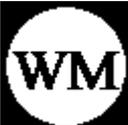<br>BER = 0<br>NCC = 1 | 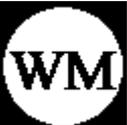<br>BER = 0<br>NCC = 1 |
| High | 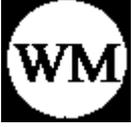 | 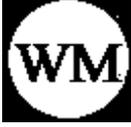 | 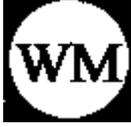 | 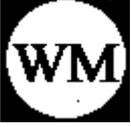 | 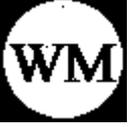 |
| | BER = 0.0002<br>NCC = 0.9998 | BER = 0.0005<br>NCC = 0.9996 | BER = 0.0009<br>NCC = 0.9991 | BER = 0.0005<br>NCC = 0.9996 | BER = 0.0007<br>NCC = 0.9993 |
| Medium | 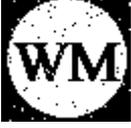 | 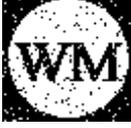 | 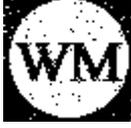 | 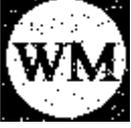 | 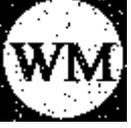 |
| | BER = 0.0173<br>NCC = 0.9841 | BER = 0.0439<br>NCC = 0.6596 | BER = 0.0251<br>NCC = 0.9769 | BER = 0.0231<br>NCC = 0.9787 | BER = 0.0232<br>NCC = 0.9787 |
| Low | 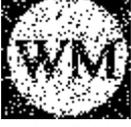 | 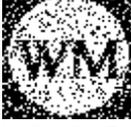 | 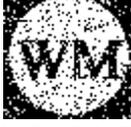 | 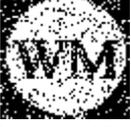 | 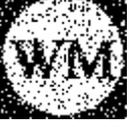 |
| | BER = 0.1104<br>NCC = 0.8970 | BER = 0.1543<br>NCC = 0.8580 | BER = 0.1174<br>NCC = 0.8911 | BER = 0.1228<br>NCC = 0.8861 | BER = 0.1129<br>NCC = 0.8952 |
| Minimum | 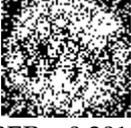 | 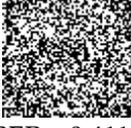 | 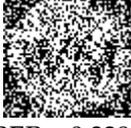 | 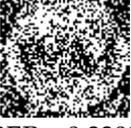 | 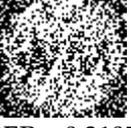 |
| | BER = 0.3018<br>NCC = 0.7177 | BER = 0.4116<br>NCC = 0.6078 | BER = 0.3293<br>NCC = 0.6902 | BER = 0.3303<br>NCC = 0.6832 | BER = 0.3105<br>NCC = 0.7142 |

**Table 2.** Experimental results for illustrations watermarked by the algorithm [19]

| Compression quality | Work1 | Email | Gears | Social media | Work2 |
|---|---|---|---|---|---|
| No | 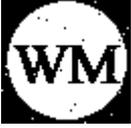 | 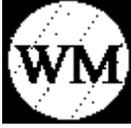 | 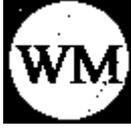 | 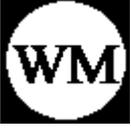 | 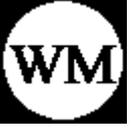 |
| | BER = 0.0054<br>NCC = 0.9951 | BER = 0.0081<br>NCC = 0.9926 | BER = 0.0034<br>NCC = 0.9968 | BER = 0<br>NCC = 1 | BER = 0<br>NCC = 1 |
| Maximum | 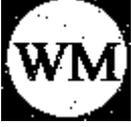 | 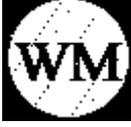 | 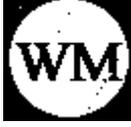 | 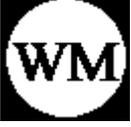 | 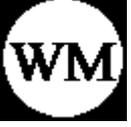 |
| | BER = 0.0068<br>NCC = 0.9937 | BER = 0.0088<br>NCC = 0.9918 | BER = 0.0034<br>NCC = 0.9968 | BER = 0<br>NCC = 1 | BER = 0<br>NCC = 1 |
| High | 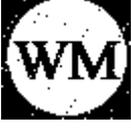 | 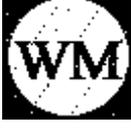 | 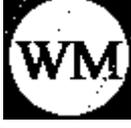 | 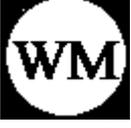 | 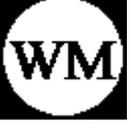 |
| | BER = 0.0115<br>NCC = 0.9895 | BER = 0.0102<br>NCC = 0.9905 | BER = 0.0042<br>NCC = 0.9962 | BER = 0.0002<br>NCC = 0.9997 | BER = 0.0002<br>NCC = 0.9997 |
| Medium | 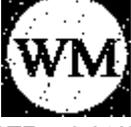 | 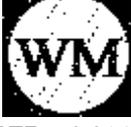 | 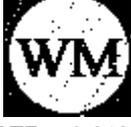 | 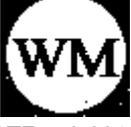 | 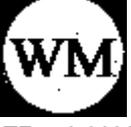 |
| | BER = 0.0187<br>NCC = 0.9828 | BER = 0.0163<br>NCC = 0.9849 | BER = 0.0139<br>NCC = 0.9872 | BER = 0.0019<br>NCC = 0.9982 | BER = 0.0022<br>NCC = 0.9979 |
| Low | 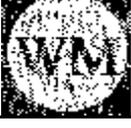 | 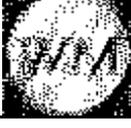 | 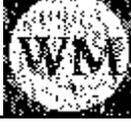 | 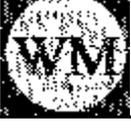 | 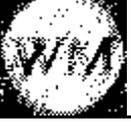 |
| | BER = 0.1359 | BER = 0.1519 | BER = 0.1313 | BER = 0.1021 | BER = 0.1174 |
| | NCC = 0.8751 | NCC = 0.8645 | NCC = 0.8792 | NCC = 0.9045 | NCC = 0.8957 |
| Minimum | 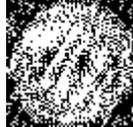 | 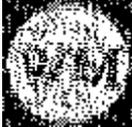 | 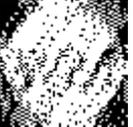 | 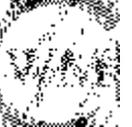 | 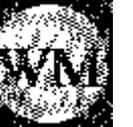 |
| | BER = 0.2851 | BER = 0.1953 | BER = 0.2456 | BER = 0.3161 | BER = 0.1716 |
| | NCC = 0.7354 | NCC = 0.8161 | NCC = 0.8028 | NCC = 0.7855 | NCC = 0.8409 |

Table 3 shows the results of the experiment with the algorithm [22]. Unlike the previous experiment, in this case the watermark is embedded in the DCT frequency domain. The algorithm is based on changing the differences in the DCT coefficients, and this makes it inappropriate for poorly detailed images, because the difference between the coefficients of adjacent blocks is often zero. Therefore, experiments on embedding the watermark in illustrations were not carried out.

Experiments show that the watermark is completely destroyed after minimum quality compression. However, when compressed at a higher quality, the watermark can be easily detected in the image. Therefore, this algorithm is also good for implementing effective electronic document protection.

**Table 3.** Experimental results for highly detailed images watermarked by the algorithm [22]

| Compression quality | Airplane | Baboon | Barbara | Goldhill | Peppers |
|---|---|---|---|---|---|
| No | 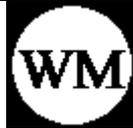 | 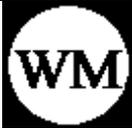 | 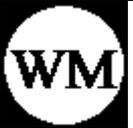 | 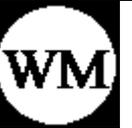 | 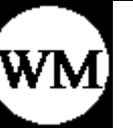 |
| | BER = 0 | BER = 0 | BER = 0 | BER = 0 | BER = 0 |
| | NCC = 1 | NCC = 1 | NCC = 1 | NCC = 1 | NCC = 1 |
| Maximum | 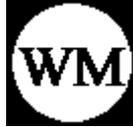 | 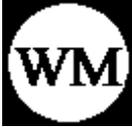 | 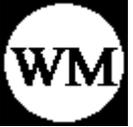 | 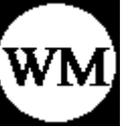 | 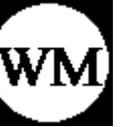 |
| | BER = 0 | BER = 0 | BER = 0 | BER = 0 | BER = 0 |
| | NCC = 1 | NCC = 1 | NCC = 1 | NCC = 1 | NCC = 1 |
| High | 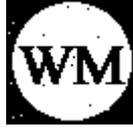 | 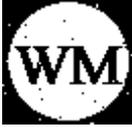 | 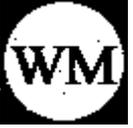 | 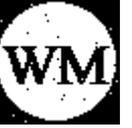 | 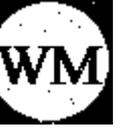 |
| | BER = 0.0044 | BER = 0.0032 | BER = 0.0022 | BER = 0.0068 | BER = 0.0046 |
| | NCC = 0.9959 | NCC = 0.9942 | NCC = 0.9979 | NCC = 0.9937 | NCC = 0.9957 |
| Medium | 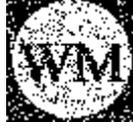 | 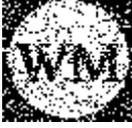 | 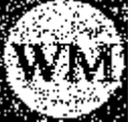 | 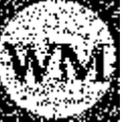 | 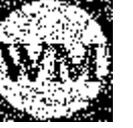 |
| | BER = 0.1243 | BER = 0.1445 | BER = 0.1289 | BER = 0.1406 | BER = 0.1421 |
| | NCC = 0.8842 | NCC = 0.8661 | NCC = 0.8799 | NCC = 0.8686 | NCC = 0.8677 |
| Low | 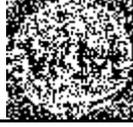 | 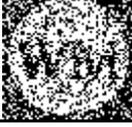 | 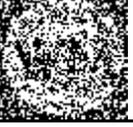 | 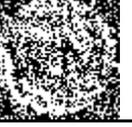 | 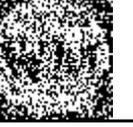 |
| Minimum | BER = 0.3129<br>NCC = 0.6976<br>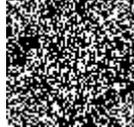<br>BER = 0.5398<br>NCC = 0.4235 | BER = 0.2605<br>NCC = 0.7552<br>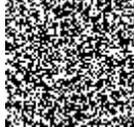<br>BER = 0.5259<br>NCC = 0.4835 | BER = 0.3196<br>NCC = 0.6964<br>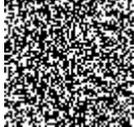<br>BER = 0.5369<br>NCC = 0.4534 | BER = 0.3113<br>NCC = 0.7036<br>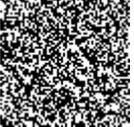<br>BER = 0.5376<br>NCC = 0.4623 | BER = 0.3142<br>NCC = 0.6977<br>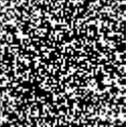<br>BER = 0.5374<br>NCC = 0.4322 |

Table 4 shows the results of the experiment using the algorithm [24]. This hybrid algorithm is based on a combination of DCT and DWT. It differs from algorithms [19] and [22] by non-blind watermark extraction. This means that the cover image is required for watermark extraction. This is not convenient in all cases, which is a disadvantage. However, experimental results show improved compression resistance. Even with minimum compression quality, the watermark can be easily detected in the image. Therefore, this algorithm can also be used for effective protection of electronic documents containing images. It is worth noting that when watermarks are embedded in illustrations, noticeable distortions of the extracted watermark occur even at maximum compression quality. According to this we can conclude that the algorithm [24] is suitable only for highly detailed images, as well as the algorithm [22].

**Table 4.** Experimental results for highly detailed images watermarked by the algorithm [24]

| Compression quality | Airplane | Baboon | Barbara | Goldhill | Peppers |
|---|---|---|---|---|---|
| No | 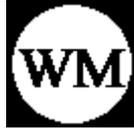 | 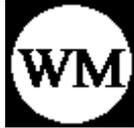 | 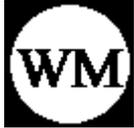 | 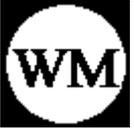 | 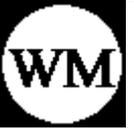 |
| | BER = 0<br>NCC = 1 | BER = 0<br>NCC = 1 | BER = 0<br>NCC = 1 | BER = 0<br>NCC = 1 | BER = 0<br>NCC = 1 |
| Maximum | 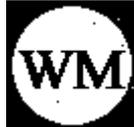 | 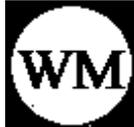 | 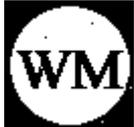 | 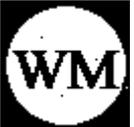 | 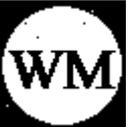 |
| | BER = 0.0012<br>NCC = 0.9989 | BER = 0.0005<br>NCC = 0.9996 | BER = 0.0012<br>NCC = 0.9989 | BER = 0.0009<br>NCC = 0.9991 | BER = 0.0012<br>NCC = 0.9989 |
| High | 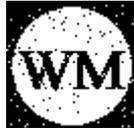 | 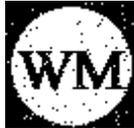 | 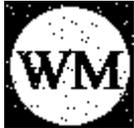 | 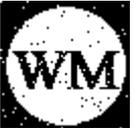 | 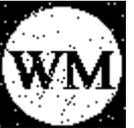 |
| | BER = 0.0261<br>NCC = 0.9760 | BER = 0.0209<br>NCC = 0.9807 | BER = 0.0168<br>NCC = 0.9845 | BER = 0.0186<br>NCC = 0.9830 | BER = 0.0227<br>NCC = 0.9791 |
| Medium | 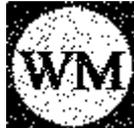 | 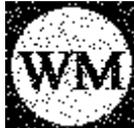 | 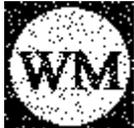 | 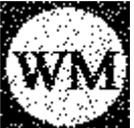 | 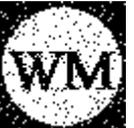 |
| | BER = 0.0537<br>NCC = 0.9504 | BER = 0.0601<br>NCC = 0.9445 | BER = 0.0637<br>NCC = 0.9412 | BER = 0.0674<br>NCC = 0.9381 | BER = 0.0588<br>NCC = 0.9456 |
| Low | 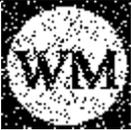 | 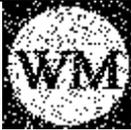 | 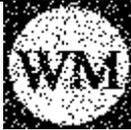 | 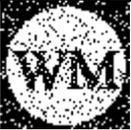 | 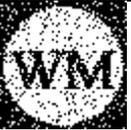 |
| | BER = 0.1047 NCC = 0.9027 | BER = 0.1111 NCC = 0.8967 | BER = 0.1082 NCC = 0.8994 | BER = 0.1089 NCC = 0.8993 | BER = 0.1089 NCC = 0.8993 |
| Minimum | 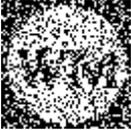 | 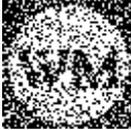 | 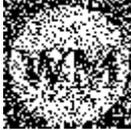 | 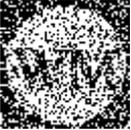 | 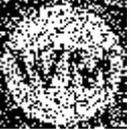 |
| | BER = 0.2522 NCC = 0.7652 | BER = 0.2488 NCC = 0.7672 | BER = 0.2627 NCC = 0.7535 | BER = 0.2515 NCC = 0.7627 | BER = 0.2456 NCC = 0.7688 |

## 5. Discussion

The experiments presented in Section 4 are carried out to confirm the practical value of the proposed technology for protecting electronic documents. Another purpose is to evaluate the effectiveness of various embedding algorithms that can be used to implement this technology.

For the experiments, algorithms are used that are representatives of different classes. This is a spatial domain embedding algorithm [19], frequency domain embedding algorithms [22] and [24]. Algorithm [22] is based on DCT. The algorithm [24] is based on DCT and DWT and is distinguished by non-blind watermark extraction. The choice of these algorithms from the set of possible ones is determined by the high embedding efficiency, which is declared by the authors of the studies. We do not use DFT-based algorithms to conduct experiments. Such algorithms provide good resistance to geometric attacks. This is due to the properties of the DFT. However, JPEG compression does not belong to this class of attacks. Therefore, DFT-based algorithms are less useful for our technology than other frequency algorithms.

Experimental results show that none of the algorithms has an absolute advantage over the rest of the algorithms. A comparison of these algorithms in terms of their resistance to converting documents to PDF is shown in Table 5.

**Table 5.** Comparison of the algorithms compression resistance

| Algorithm | Highly detailed images | Illustrations |
|---|---|---|
| [19] | • Noticeable distortions of the digital watermark appear when the quality of the compressed images is low.<br>• The watermark is almost destroyed when the quality of the compressed images is minimal.<br>• The pattern of digital watermark distortion is the same for different images | • Noticeable distortions of the digital watermark appear when the quality of the compressed images is low.<br>• The watermark is significantly distorted but not destroyed when the quality of the compressed images is minimal.<br>• The pattern of the distortion of a digital watermark is noticeably different for different images |
| [22] | • Noticeable distortions of the digital watermark appear when the quality of the compressed images is medium. | Embedding is not carried out |
| [24] | <ul><li>The watermark is almost destroyed when the quality of the compressed images is low.</li><li>The watermark is completely destroyed when the quality of the compressed images is minimal.</li><li>The pattern of digital watermark distortion is the same for different images</li><li>Noticeable distortions of the digital watermark appear when the quality of the compressed images is low.</li><li>The watermark is significantly distorted but not destroyed when the quality of the compressed images is minimal.</li><li>The pattern of digital watermark distortion is the same for different images</li></ul> | Embedding is not carried out |

The advantage of the spatial domain embedding algorithm is that it is applicable to both highly detailed images and poorly detailed illustrations. Moreover, in the case of digital watermarks embedding in illustrations, this algorithm shows the great resistance to compression.

Frequency domain algorithms are poorly applicable to illustrations. Therefore, we can conclude that embedding into such images should be carried out using spatial domain algorithms. Special frequency algorithms can also be developed for this purpose.

When embedding digital watermarks into highly detailed images, the best result is shown by a non-blind algorithm based on a combination of DCT and DWT. This advantage is due to the use of the original image when extracting the digital watermark.

The experiment results may lead to the conclusion that the use of algorithms of one class within the proposed electronic documents protection technology is not advisable. The good solution is to use them together, depending on the characteristics of a particular electronic document.

Therefore, we propose the following scenario for the implementation of our technology:

1. To embed digital watermarks into images in electronic documents, a pool of algorithms is formed that covers various groups of images.

2. The analysis and classification of the images contained in the document are performed before embedding. The level of image detail is used as a classification criterion.

3. Digital watermarks are generated with or without the context of the document.

4. The generated digital watermarks are embedded into images in the document. The choice of the embedding algorithm for each image is performed based on the image class.

## 6. Conclusion

The paper presents a new technology for protecting the authorship of electronic documents. To ensure authorship protection, it is proposed to embed invisible watermarks in the images contained in the document. The key feature of the proposed technology is that the content and structure of the document do not change in the watermarking process. Our technology is useful for different classes of documents containing images.

An analysis of the proposed technology applicability shows that it effectively provides proof of authorship both when copying an entire document, and when copying parts of it. To ensure a high level of protection, it is recommended to use robust watermarking algorithms. It is advisable to use a combination of spatial and frequency domain embedding algorithms, depending on the level of image detail.

**Conflict of interest**

None.

**Acknowledgements**

This research was funded by the Ministry of Science and Higher Education of Russia, Government Order for 2020–2022, project no. FEWM-2020-0037 (TUSUR).